# Direct Band Gap Semiconducting Holey Graphyne: Structure, Synthesis and Potential Applications

Xinghui Liu[1,2*], Soo Min Cho[2*], Shiru Lin[3], Eunbhin Yun[1,2], Eun Hee Baek[2], Zhongfang Chen[3], Do Hyun Ryu[2‡], Hyoyoung Lee[1,2,3‡]

1 Centre for Integrated Nanostructure Physics (CINAP), Institute of Basic Science (IBS), 2066 Seoburo, Jangan-Gu, Suwon 16419, Republic of Korea.

2 Department of Chemistry, Sungkyunkwan University (SKKU), 2066 Seoburo, Jangan-Gu, Suwon 16419, Republic of Korea.

3 Department of Chemistry University of Puerto Rico, Rio Piedras Campus, San Juan, USA Campus

[*]These authors contributed equally to this work.

[‡]Corresponding author. Email: dhryu@skku.edu (D.H.R.), hyoyoung@skku.edu (H.L.)

**Abstract:**

Here we report two-dimensional (2D) single-crystalline holey-graphyne (HGY) created an interfacial two-solvent system through a Castro-Stephens coupling reaction from 1,3,5-tribromo-2,4,6-triethynylbenzene. HGY is a new type of 2D carbon allotrope whose structure is comprised of a pattern of six-vertex and eight-vertex rings. The carbon-carbon 2D network of HGY is alternately linked between benzene rings and *sp* (carbon-carbon triple bond) bonding. The ratio of the *sp* over *sp²* bonding is 50%. It is confirmed that HGY is stable by DFT calculation. The vibrational, optic, and electric properties of HGY are investigated theoretically and experimentally. It is a p-type semiconductor that embraces a natural direct band gap (~ 1.0 eV) with high hole mobility ($2.13 \times 10^9$ cm$^2$ V$^{-1}$ s$^{-1}$) and electron mobility (~ $10^4$ cm$^2$ V$^{-1}$ s$^{-1}$) at room temperature. This report is expected to help develop a new types of carbon-based semiconductor devices with high mobility.

Tremendous efforts have been invested in research in the area of two-dimensional (2D) periodicity semiconducting devices. Recently, novel carbon allotropes such as fullerene,(*1*) carbon nanotube,(*2*) graphene,(*3*) graphyne(***4-8***) and graphdiyne(*4-7, 9*) have been explored, due to their outstanding properties. Graphene research have made significant advances in modern chemistry and physics because of its fascinating properties(***10***). However, the electronic structure of graphene has a limiting property for the use of semiconducting materials. Therefore, it is necessary to find a new type of 2D carbon allotropes that have exceptional semiconducting properties, such as band gap and high mobility. Herein, we report novel holey-graphyne (HGY), which is a new type of 2D carbon allotrope whose structure is comprised of a pattern of six-vertex and eight-vertex rings hybridized

Firstly, we investigated the structure information of HGY and its stability in terms of kinetic, mechanical and thermodynamics aspects. It shows an excellent stability, which are essential properties for the new 2D crystal (Fig. S8-9 and Supporting Text 1). For experimentally synthesizing HGY, we designed new synthetic route using Castro-Stephens coupling reaction(*11*) (Fig. 1a). For this reaction, a new starting monomer, 1,3,5-tribromo-2,4,6-triethynylbenzene **1** was prepared. As shown in Fig. 1a, the known 2,4,6-tribromobenzene-1,3,5-tricarbaldehyde **4** was prepared from mesitylene by modifying reported procedure(*12*) with an overall yield of 41% in a total of 4 steps. The 1,3,5-tribromo-2,4,6-trimethylbenzene was reacted with bromine under irradiation of light. While the reported TEMPO oxidation of (2,4,6-tribromobenzene-1,3,5-

triyl)trimethanol **3** afforded **4** in 20% yield, we used PCC oxidation with celite and molecular sieves to obtain the intermediate **4** showing in 83% yield. Finally, the monomer **1** was converted from the trialdehyde **4** using Corey-Fuchs reaction(*13*) with lithium diisopropylamide as base. The detailed reaction procedures (Figs S1-5), characterizations including $^1$H NMR and $^{13}$C NMR spectra (Figs S6-7), and reaction yields are shown in experimental section of supporting information. Following synthetic method shown in Fig. 1b, the monomer **1** was transformed to the crystalline HGY film on interface between organic phase and water phase *via* 'Biradical' intermediate. The synthesized HGY was identified using $^{13}$C-solid NMR. The $^{13}$C-solid NMR (Fig. S15b) shows 4 main peaks. The two peaks at 91.1 and 80.9 ppm are assigned to the sp carbon atoms whereas the two peaks at 129.7 and 126.8 ppm are attributed to the sp$^2$ carbon.

The morphologies of the HGY film were characterized by optical microscopy (OM) and scanning electron microscopy (SEM), as shown in Fig. 2a. The HGY film, with the lateral dimension up to several micrometers, is shown in Fig. 2a. Furthermore, the film was a flat sheet with a thickness of 5.3 nm by atomic force microscopy (AFM) characterization of HGY transferred on the Si/SiO$_2$ substrate (Fig. 2b). The 3D image of the HGY film is shown in Fig. S10. Transmission electron microscopy (TEM) was used to analyze the morphology and microstructure of the film. Fig. S11 and 2c display the low-resolution TEM image and the high-resolution TEM image, respectively, of the film, and further verify the successful synthesis of continuous film. The inset (Fig. 2c) shows the corresponding the fast Fourier transform (FFT) pattern of the HRTEM image, indicating the single crystalline line nature of the film, which means that the structure of the film shows significant in-plane order over a relatively large area, and exhibits the single-crystal diffractions spots. An enlarged inverse FFT (IFFT) image of the area highlighted by the red square of Fig. 2c is shown in Fig. 2d, which indicated the d-spacing is 0.442 nm, showing an index as (200) reflections of HGY. Another set of hexagonally arranged diffraction spots with a larger d-spacing of 0.75 nm was observed, which can be indexed by the (220) reflections of HGY. In addition, the crystal structure of the HGY was examined by X-ray diffraction (XRD) measurement. As illustrated in Fig. 2e, the (002) peak at approximately 24.5° for HGY indicates an interlayer spacing of 0.36 nm. Moreover, lattice fringe image of HRTEM (Fig. 2f and 2g) reveals curve streaks with a lattice parameter of 0.36 nm, in good agreement with the XRD result of the (002) peak, which can be assigned to the spacing between carbon layers (Fig. 2h). Fig. 2g showing the line profiles from the blue line indicate that the interlayer spacing between HGY layers is 3.60 nm. This value is slightly higher than that of graphene,(*14*) owing to the more delocalized system(*15, 16*) of HGY.

Different stacking modes result in significant variations in electronic and optic properties as compared to single layer 2D materials(*17*). Here, the most stable mode of HGY is proved to be the ACB stacking mode. The side view and top view of ACB stacking modes are displayed in Figs. 2h and 2i, respectively. There have been various stacking modes to calculate the binding energies and band structures of (bi) tri-layer HGY using van der Waals corrected DFT (Fig. S12, Table S1 Supporting Text 2).  The elemental composition and bonding structure were probed systemically with energy-dispersive X-ray spectroscopy (EDS), X-ray photoelectron spectroscopy (XPS), Fourier-transform infrared spectroscopy (FT-IR) and Raman spectroscopy. The survey spectra of XPS (Fig 2a and S13) indicate that the HGY is composed primarily of elemental carbon, with silicon and oxygen signals from the underlying $SiO_2$ layer. The corresponding STEM-EDS maps (Fig. 3b) show that the distributions of C, Br and O and atomic percentage of carbon and oxygen are 96.69% and 3.31%, respectively. Any peak of bromine is almost not found in EDS spectra and carbon is chiefly located on the film, which proves the completion of Castro-Stephens coupling reaction. The electron image and EDS analysis are shown in Fig. S14, which is identical to the result of XPS. High-resolution XPS can discriminate the circumstance of an element. The C 1s peak at 284.8 eV is deconvoluted into four Gaussian sub peaks (Fig. 3a), which are mainly contributions from *sp²* (C=C) at 284.5 eV and *sp* (C≡C) at 285.3 eV species. In addition, the abundance ratio of the *sp/sp²* carbon is 1.0, which is in good agreement with the chemical composition of the HGY structure

The sub peaks C-O and C=O, with minor contributions to C 1s peak, are located at 286.6 eV and 288.5 eV, respectively. The presence of elemental O derives from the absorption of air in the pores and the oxidation of terminal alkyne, which will cause some defects. Meanwhile, the HGY and monomer **1** were also analyzed by FT-IR spectroscopy (Fig. 3c). In each measurement, the signal was normalized by the highest intensity peak. The C-Br stretching at 682 cm$^{-1}$ and the C≡C-H stretching at 3278 cm$^{-1}$ in monomer **1** decrease when the 2D crystalline HGY were formed, separately. The peak C≡C stretching modes (Figs. 3c and S15a), with zero dipole moment, were identified at 1933 cm$^{-1}$ and 2115 cm$^{-1}$, respectively. Meanwhile, C-H stretching modes show bands at 2852 cm$^{-1}$ and 2921 cm$^{-1}$, respectively. The strong stretching of the carbon rings is shown at 1336 cm$^{-1}$, indicating the $E_{1u}$ symmetry mode of HGY, which clearly confirms the molecular HGY structure.

Raman scattering spectroscopy is a powerful instrument, providing the ability to analyze the structural information of HGY, particularly for those Raman-active alkyne linkers arrayed in the concrete topology of HGY. We experimentally and theoretically investigated the novel structure of HGY, where we studied by group theory and density functional theory (DFT) calculations. HGY, like graphene, has the same hexagonal symmetry as the point group $D_{6h}$ (No. 27) in the Schönflies notation and space group *P6/mmm* (No. 191) in the Hermann-Mauguin notation. The Raman scattering process is limited to phonons at Gamma point in the Brillouin zone (BZ) center, due to the law of momentum conservation. The irreducible representations(*18*) of all modes at the center of BZ are described in Fig. S16. Specifically, the irreducible representation of acoustic modes, the irreducible representation of optic modes, and the Raman tensors of the active modes are shown in Fig. S16. All the symmetry properties of vibrational modes with the Raman-active are shown in Fig. S17. Experimentally, all Raman spectra were in the range of 200 cm$^{-1}$ to 2500 cm$^{-1}$, which agreed with the investigation of DFT (all the peaks were normalized by one unit and a half unit after and before breakpoint, respectively, in Fig. 3d). The two prominent peaks of C≡C (*sp*) at 1946 cm$^{-1}$ and 2207 cm$^{-1}$ were derived from the $A_{1g}$ and $E_{2g}$ vibration modes of the conjugated alkyne linkage, respectively, which was also consistent with the two stretching modes at 1933 cm$^{-1}$ and 2115 cm$^{-1}$ in FT-IR, as shown in Fig. S15. Terminal alkyne vibrational modes (2100-2120 cm$^{-1}$) were not observed at all, indicating that all terminal alkyne residues of monomers were formed with π-conjugated network structures(*19*), which agreed with the EDS data in Fig. 3b and Fig. S14. The peak at 1525 cm$^{-1}$ corresponds to the first-order scattering of the $E_{2g}$ vibration mode inspected for in-phase stretching of *sp²* carbon, which is red shifted compared to the G band of graphite (1575 cm$^{-1}$)(*20*). The peak at 1296 cm$^{-1}$ corresponds to the breathing vibration of *sp²* carbon domains in aromatic rings (D band). The peak at 950 cm$^{-1}$ comes primarily from the breathing vibration of the benzene ring; the peak at 804 cm$^{-1}$ can be attributed to the breathing vibration between alkyne-related rings; the peaks at 682 cm$^{-1}$, 575 cm$^{-1}$, and 294 cm$^{-1}$ can be classified as the vibration of $E_{1g}$ modes between the layers; the peak at 474 cm$^{-1}$ can be ascribed to the scissoring vibration of the atoms in alkyne-related rings.

Then, we investigate the electronic structure and project density of states (PDOS) of HGY. The three-dimensional energy band structure is shown in Fig. 4a. The conduction band and the valence band, two Dirac cones, have two equal vertices K and K' = -K points in the hexagonal Brillouin zone, with a similar valley symmetry to graphene(*21*). In addition, the energy spectrum is electron-hole symmetric (Figs. 4a and b), where the scattering mechanism and transport properties are identical for electron and hole doping cases. Furthermore, the band dispersion arises mostly from the overlap of the carbon $2p_z$ orbitals when we combine the band structure and PDOS (Fig. 4c). It is clearly seen that HGY is a semiconductor with a direct gap of 0.5 and 1 eV at the K-point, calculated at level of PBE and HSE06 functional theory, separately. It should be noted that the band gap of semiconducting materials is underestimated by the PBE method. The band structures and PDOS are displayed in Fig. S18, which clearly shows the difference between the PBE and HSE06 functions. The wave functions of the valence band maximum and the conduction band minimum are discussed in Fig. 4d and Supporting Text 3. For the fabrication of electronics devices or the maximization of an energy conversion efficiency in solar cells, the work function (Φ) is an important parameter. It can be defined as the energy required to remove an electron from the highest filled level in the Fermi distribution of a solid to vacuum at absolute zero Kevin. The work function of HGY is 5.2 eV, according to the DFT calculations shown in Fig. S19.

The acoustic phonon-limited mobility in HGY monolayer was examined, which is essential in nanoelectronics engineering. Fig. 4e shows the first Brillouin zone (FBZ), which is related to the hexagonal (red line) and orthogonal (blue line) lattice. The K point (fractional k-space coordinates (-1/3, 2/3, 0) defined in the reciprocal lattice of the primitive hexagonal cell is folded into (0, 1/3, 0) point sitting at a direction in the FBZ of the orthogonal supercell in Fig. 4e. Compared to the rhombus drawn with dashed lines in Fig. 4f, there are 48 atoms in this rectangle drawn with dashed lines (which has lattice constants $a_0$ = 18.76 Å and $b_0$ =10.83 Å) to show the orthogonal supercell at equilibrium geometry with the primitive cell. In this charge transport, the orthogonal supercell along two vertical direction x and y for HGY sheet which can more instinctually be explained for the transport property as shown in Fig. 4g. First, the effective mass (m*) values for both electron and hole along the x and y directions are 0.15 and 0.21 $m_e$, respectively ($m_e$ is the free electron mass), and can be calculated from the energy dispersion equation: m* = $\hbar^2([\partial^2\varepsilon(k)/\partial k_x \partial k_y]^{-1}$. Then, the elastic moduli (C) can be computed by performing lattice analysis and deriving the elastic constants from the strain-stress relationship,(22) which includes the contributions from distortions with rigid ions and the ionic relaxations, as shown in Table 1. Finally, the deformation potential ($E_1$) can be determined by calculating band edge positions of VBM and CBM as functions of dilating the lattice constant. The shift of the band edges as functions of strain along x and y directions are shown in Figs. S20b-c, respectively. Through dilation of the lattice along the x and y directions, the DP constant can be calculated as $E_1$= $\partial E_{edge}/(\partial \frac{a}{a_0})$ along the x direction and $E_1$= $\partial E_{edge}/(\partial \frac{b}{b_0})$ along the y direction.

To better understand the charge transport behaviors of HGY under the effective mass approximation and the electron-acoustic phonon coupling mechanism, Bardeen and Shockley derived an analytical expression for the intrinsic carrier mobility(23), which have been extensively applied to study μ in 2D(24-26) materials with the following equation:

$$\mu_{2D} = \frac{\tau e}{|m^*|} = \frac{2e\hbar^3 C}{3k_B T |m^*|^2 E_1^2}$$　　　　　　　　Eq. (1)

Based on the obtained $E_1$, C, and m* using Eq. 1, the acoustic phonon-limited mobility at room temperature (300K) and the relaxation time are calculated. The results are displayed in Table 1.

The obtained intrinsic electron mobility is 9.97 ×10$^4$ cm$^2$ V$^{-1}$ s$^{-1}$ along the x direction and 4.83 × 10$^4$ cm$^2$ V$^{-1}$ s$^{-1}$ along the y direction at room temperature (RT). However, the intrinsic hole mobility is approximately 5 orders of magnitude higher than the mobility of the electrons, which is 2.13 × 10$^9$ cm$^2$ V$^{-1}$ s$^{-1}$ along the x direction and 1.38 × 10$^9$ cm$^2$ V$^{-1}$ s$^{-1}$ along the y direction. The acoustic phonon scattering relaxation times are also computed Eq. 1 after we got the mobility. They are 8.49 ps (in the x direction) and 5.76 ps (in the y direction) for electrons. These values are similar to the corresponding values for graphene and graphdiyne.(24, 27) But the relaxation time of holes is also five times larger than that of electron which is 0.18 and 0.16 μs along x and y directions, respectively. Generally, graphene has high mobility due to the Dirac cone at the K point, which is the result of massless of electron and hole, which also attributes to the high mobility of HGY. It is important to understand why HGY has a small deformation potential (DP) constant. The DP constant can characterize the coupling capability between electron or hole with the acoustic phonon (Fig.S20 and Supporting Text 4). For the fabrication of the field effect transistor (FET) devices, all fabrication steps should be carried out in clean room and can be probed on clean and well-defined substrates, which will be reported elsewhere in the near future.

**Figures**

**(a)**

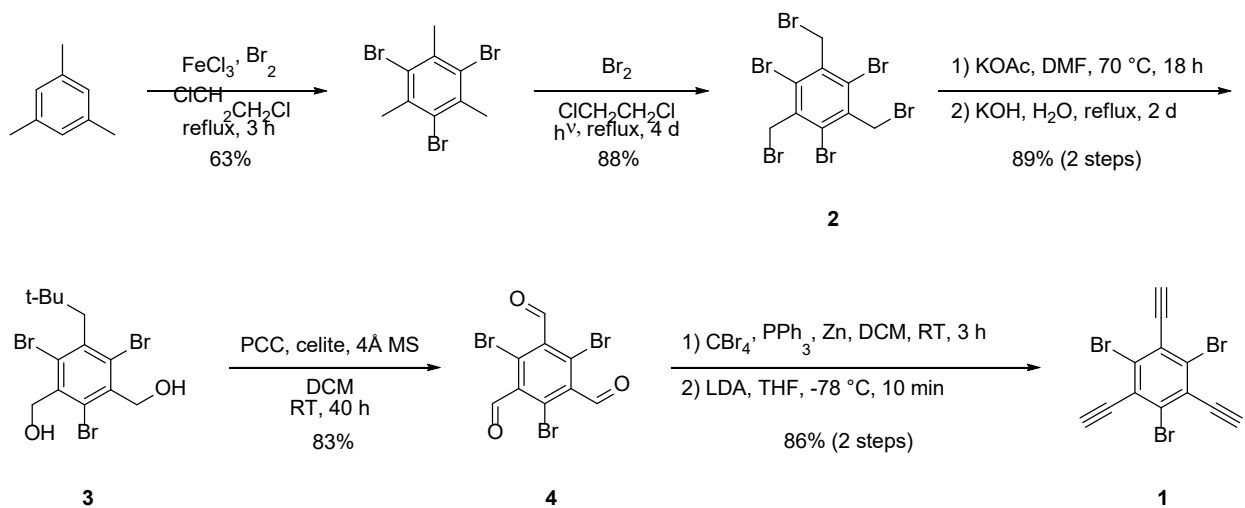

**(b)**

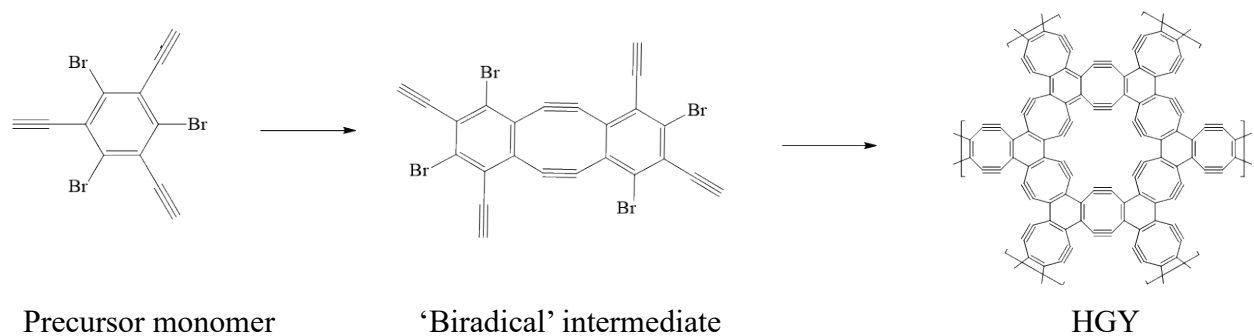

Precursor monomer      'Biradical' intermediate      HGY

**Figure 1. Bottom-up fabrication of atomically precise HGY.** (a) Schematic of proposed structure of synthesis of precursor monomer 1,3,5-tribromo-2,4,6-triethynylbenzene (**1**). (b) Schematic of the formation process from **1** to HGY using Castro-Stephens coupling mechanism.

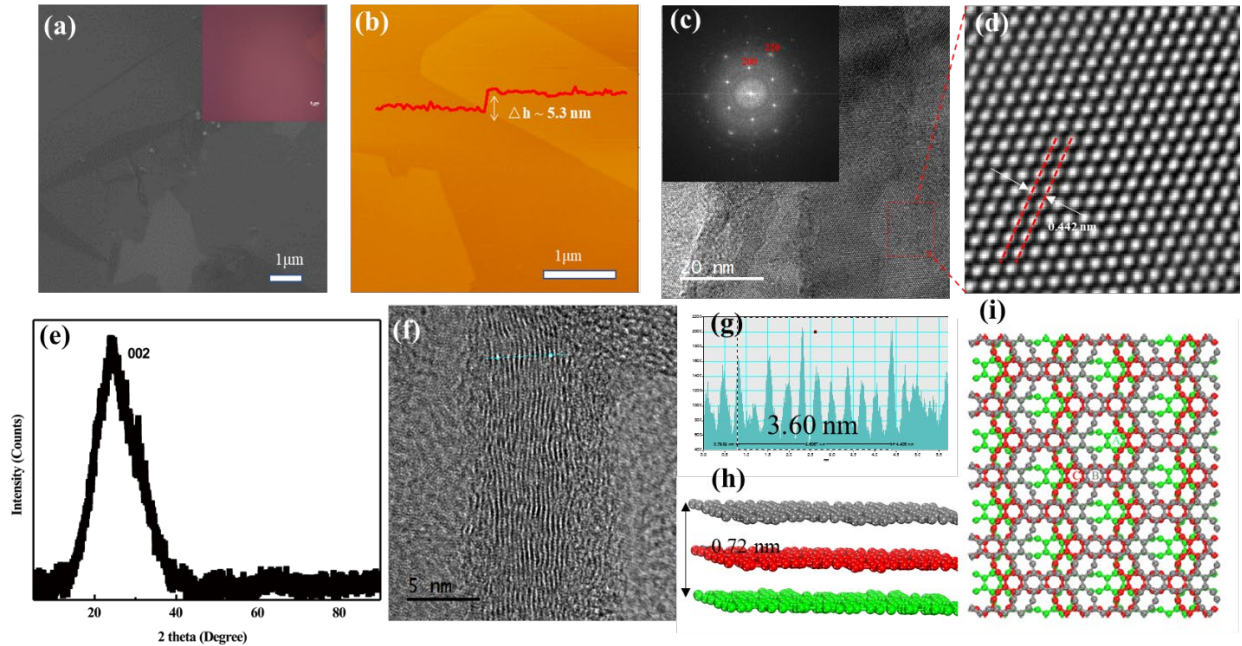

**Figure 2. The morphologies and structure characterization of HGY film.** (a) SEM image and optical microscopy (OM) image (inset) of HGY. (b) AFM image of HGY on Si/SiO$_2$ substrate. (c) High resolution TEM (HRTEM) image of HGY. Related Fourier transform results (FFT) patterns are in corresponding insets. (d) Enlarged Inverse FFT (IFFT) image of the area highlighted by the red square. (e) The crystal structure of the HGY examined by X-ray diffraction (XRD). (f) The lattice fringe of an HRTEM image of HGY that reveals curve streaks with lattice parameter of 0.36 nm and agreement with the XRD pattern shown in Figure 2e. (g) The line profiles from the blue line indicate that the interlayer spacing between HGY layers is 0.36 nm. (h and i) The most stable modes from the simulations is ACB stacking of tri-layer HGY, which is corresponding to the side view and top view, respectively.

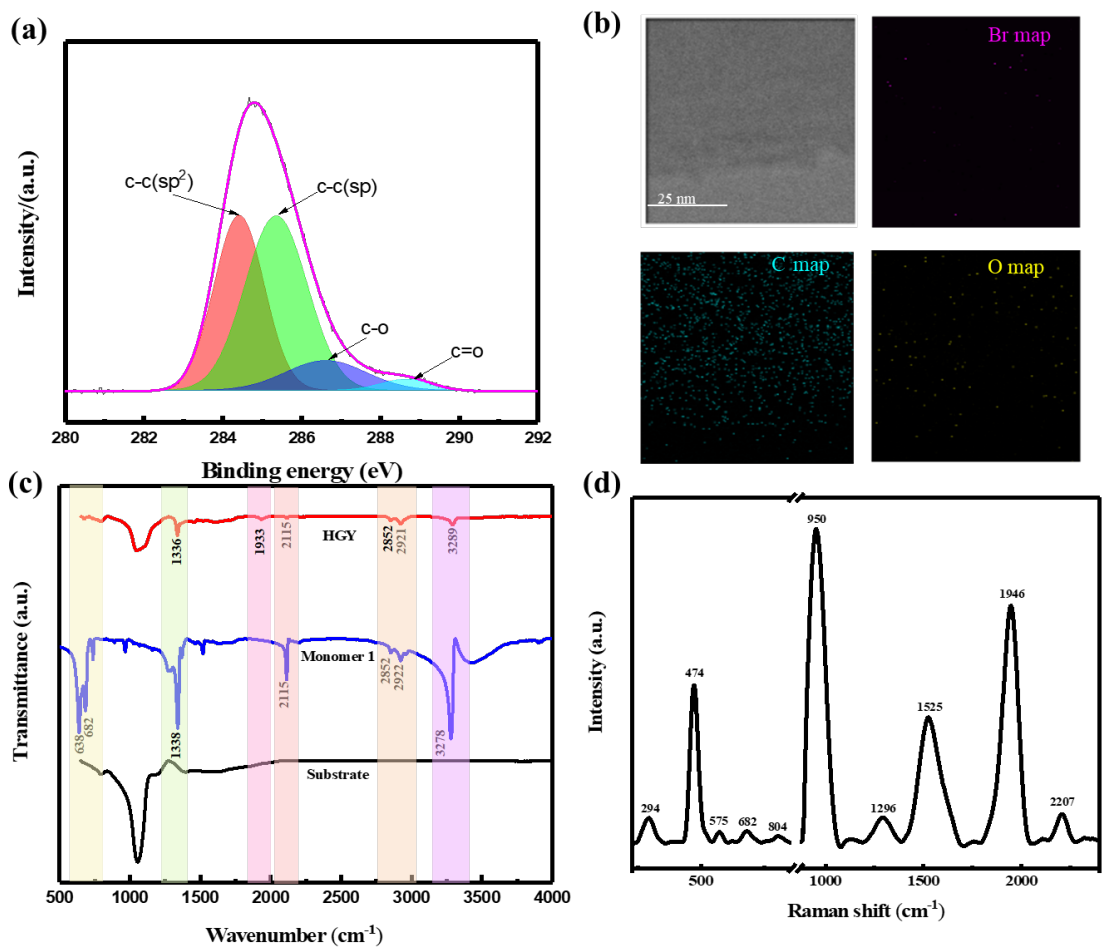

**Figure 3. Spectroscopic characterization of HGY.** (a) High-resolution core-level XPS spectrum of C 1s of HGY indicating the *sp*/*sp$^2$* ratio was 1:1. (b) STEM-EDS element mapping images of HGY showing that the HGY film, mainly composed of elemental carbon. (c) FT-IR spectra of HGY, Monomer **1,** and substrate (all data are normalized). The big difference between the monomer **1** and HGY is the new peak that is observed at 1933 cm$^{-1}$, indicating C-C coupling bonding. (d) Raman spectra of HGY (all the peaks were normalized after/before breakpoint)

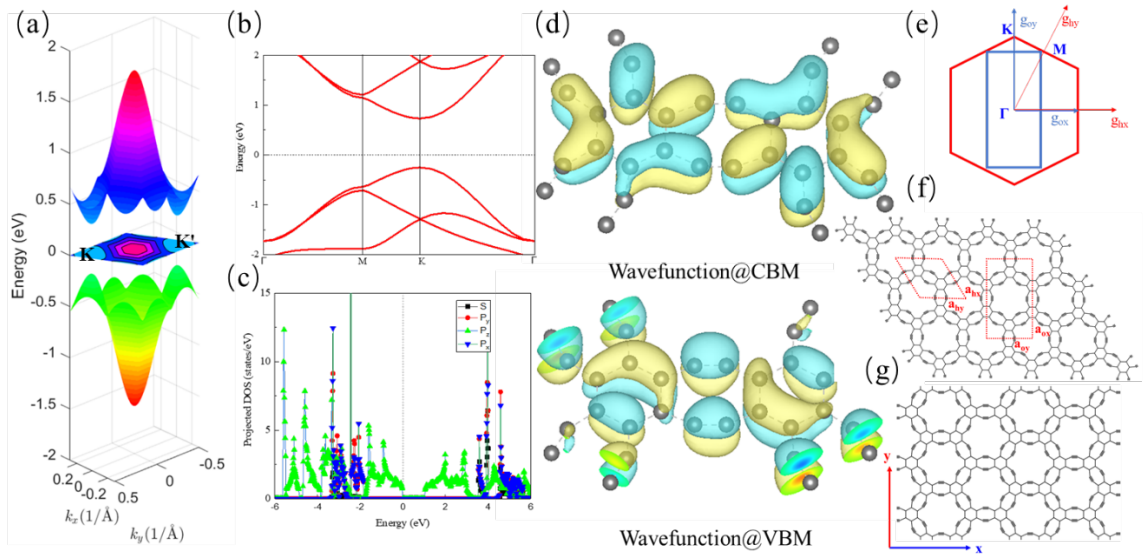

**Figure 4. Electronic property characterization of monolayer HGY sheet.** (a) Three-dimensional (3D) electronic band structure of HGY. (b) The band structures. (c) Partial density of states of HGY by HSEO6 functional. (d) K point degenerate valence band maximum wave function (Wavefunction@VBM) and conduction band minimum wave function (Wavefunction@CBM) for HGY. (e) The first Brillouin zone (FBZ) associated with the two lattices. The black dashed line shows the folding of the K point in FBZ of the hexagonal lattice into FBZ related to the orthogonal lattice. (f) Atomic structure model of monolayer HGY. The dashed lines represent the primitive hexagonal cell (defined by $a_{hx}$ and $a_{hy}$) and the orthogonal supercell (defined $a_{ox}$ and $a_{oy}$). (g) The orthogonal supercell is defined as the x and y directions for the mobility calculation.

**Table 1** The deformation potential ($E_1$), in-plane elastic constant ($C^{2D}$), effective mass ($m^*$), mobility($\mu$) and relaxation time ($\tau$) for electron (e) and hole (h) along x and y direction in 2D monolayer HGY sheet at 300 K.

| Carrier type | $E_1$(eV) | $C^{2D}$(N/m) | $m^*(m_e)$ | $\mu$ ($10^4$ cm$^2$V$^{-1}$s$^{-1}$) | $\tau$(ps) |
|---|---|---|---|---|---|
| $e^y$ | 0.7800 | 91.06 | 0.21 | 4.83 | 5.76 |
| $h^y$ | 0.0046 | 91.06 | 0.21 | 138877.73 | 165706.38 |
| $e^x$ | 0.7600 | 91.06 | 0.15 | 9.97 | 8.50 |
| $h^x$ | 0.0052 | 91.06 | 0.15 | 213008.86 | 181541.64 |

SUPPLEMENTARY MATERIALS

ACKNOWLEDGMENTS

This work was supported by the Institute for Basic Science (IBS-R011-D1) and National Research Foundation of Korea (NRF) grants funded by the Korean government (MSIP) (No.2019R1A4A2001440). We thank Prof. Zhigang Shuai and Jinyang Xi for their helpful discussion. We also thank Prof. Munseok Jeong group for the Raman measurement and discussed for the Raman data. Suar Oh took AFM measurement and Yeseul Hong took SEM measurement.